\def \pa {\partial}

\def \ln {{\rm \ ln \  }}

\def \ee {{\rm e}^}

\def \k1 {{1\overk}}

\def \a {\alpha}
\def \b {\beta}
\def \hi {\chi}
\def \l {\lambda}

\def \ep {\epsilon}
\def \r {\rho}
\def \g {\gamma}

\def \k {\kappa }
\def \d {\delta}
\def \o {\omega}
\def \s {\sigma}

\def \ph{\phi}

\def \m{\mu}
\def \n{\nu}

\def \vp {\varphi}

\def \e#1 {{\rm e}^{#1}}

\def \1p {{1\over  \pi }}
\def \2p {{{1\over  2\pi }}}
\def \4p {{ {1\over 4 \pi }}}
\def \8p {{{1\over 8 \pi }}}

\def \hg {{\hat g}}

\newcommand{\rf}[1]{(\ref{#1})}
\newcommand{\beq}{\begin{equation}}
\newcommand{\eeq}{\end{equation}}
\newcommand{\bea}{\begin{eqnarray}}
\newcommand{\eea}{\end{eqnarray}}
\newcommand{\beas}{\begin{eqnarray*}}
\newcommand{\eeas}{\end{eqnarray*}}

\documentstyle[titlepage,12pt]{article}
\begin{document}

\begin{tabbing}
\` FIT-HE-95-31\\
\` March 1995\\
\` hep-th/xxxxxx
\end{tabbing}
%%%%%%%%%%%%%%%%%%%%%%
\addtolength{\baselineskip}{0.2\baselineskip}
\begin{center}
\vspace{36pt}
  {\large \bf Quantized 2d Dilaton Gravity and \\
                      Black Holes}
\end{center}
\vspace{36pt}
\begin{center}
{\bf Kazuo Ghoroku}\footnote{e-mail: gouroku@dontaku.fit.ac.jp}
\end{center}
\vspace{2pt}
\begin{center}
Department of Physics, Fukuoka Institute of Technology,Wajiro, Higashiku,
Fukuoka 811-02, Japan
\end{center}
\vspace{36pt}
\begin{center}
{\bf abstract}
\end{center}
We have examined a modified dilaton gravity whose
action is separable into the kinetic and
the cosmological terms for the sake of the quantization.
The black hole solutions survive even in
the quantized theory, but the ADM mass of the static
solution is unbounded from below. Quantum gravitational
effect on the interacting matter field is also examined.

\newpage
%%%%%%%%%%%%%%%%%%%%%%%%%%%%%%%%  Main  %%%%%%%%%%%%%%%%%%%%%%%%%%%%%%%%%%%%

\section{Introduction}

The quantum theories of the dilaton gravity (DG) have been studied
by many authors \cite{bil}\cite{alw}\cite{gid}\cite{tan} in order to
resolve the problems related to the Hawking radiation of black holes
\cite{cal}. The necessity of the
full quantization of DG has been pointed out because of a
singularity found in a semiclassical approach based on the $1/N$
expansion \cite{ban}. The quantized model,
which has previously proposed
\cite{bil}\cite{alw}\cite{gid}, has been obtained according to the
principle of conformal invariance of the quantized theory.
The resultant effective action is written in the
form of a non-linear $\s$-model where the cosmological term is
added as a marginal operator to a simple conformally invariant theory.
The field variables in this
effective action are constructed so that they reproduce a
semi-classically quantized
dilaton gravity, in which the necessary Liouville terms are included,
at the weak coupling limit, $\ee{\phi}<<1$.
However the original DG is an interacting system of the dilaton,
so the path integral
measure for the dilaton field would not be the usual anomaly term
which is given for the conformal fields without interactions.

Here we take care of this point for the quantized measure, and a new
quantization scheme is proposed by giving a modified dilaton gravity
which is equivalent to the original DG classically.
The action of our model can be separated to two parts, the kinetic
and the cosmological terms. The kinetic part is
easily quantized according to the method of \cite{kpz}\cite{ddk}.
The cosmological term can be regarded as a perturbation for small
cosmological constant ($\m$).
Then we can use the technique used in \cite{amb} for the
quantization of this theory. Although our method
is valid for small $\m$, this is not an
obstacle to study the problems related to the black holes since
the qualitative properties of DG does not depend on the
magnitude of $\m$ as far as it is non-zero.
Here we give the quantized DG up to the $O(\m^2)$, and the qualitative
properties of the solutions of the equations of motion are
discussed. Another purpose in this paper is to examine the quantum
effects of DG on the interacting matter fields, especially on the
renormalization group equations.

Usually $N$ conformal
matter fields are added to DG. But
they are not suitable for seeing the gravitational effects
on themselves.
Previously we have shown an universal gravitational
effect on the renormalization group equations of the interacting
matter fields \cite{amb}.
Then we consider here a matter field which has its
self-interaction in order to study the same quantity.
It can be shown that the same form of
the corrections are found also in the case of DG.
But the physical scale factor, which appears in the formula,
is different from the previous case. The reason is that this scale
factor is depending on the gravitational models.
This difference is also seen in the string susceptibility
($\g_s$) of the theory.
We can see that $\g_s$ in DG for small $\m$ is real for
any number of conformal matter fields. This is also pointed out
in \cite{tan} by a semi-classical approach.

\section{A model equivalent to dilaton gravity}

Our starting point is the following DG \cite{cal}
which is written by the metric, dilaton
$\ph$ and $N$ conformal matter fields $f_i$ ($i=1\sim N$);

\beq
S_{\rm dil}=
{1 \over 4\pi}\int\,d^2z\sqrt{g} e^{-2\phi}\left\{
            \bigl[ R+4(\nabla\phi)^2+4\m^2\bigr]
                  -{1 \over 2}\sum_{i=1}^N(\nabla f_i)^2
             \right\},
\label{eq:a1}
\eeq
\noindent where $\nabla_\m$ denotes the covariant derivative with
respect to the 2d metric $g_{\m\n}$. As long as the cosmological
constant $\m$ is nonzero, this action provides non-trivial
solutions, (i) static black hole solutions and (ii) the one
which describes the formulation of a black hole by the infall
of an arbitrary pulse of massless scalar matter.

This has been extensively explored as a
toy model of four dimensional gravity
in order to make clear the problems concerned with the black holes.
Since the properties of DG mentioned
above do not depend on the the
magnitude of the parameter $\m^2$, we consider the case
of very small $\m^2$ in order to use it as a perturbation parameter.
$S_{\rm dil}$ has still a complicated form for $\phi$, so
we consider a slightly modified form, which is classically equivalent to
\rf{eq:a1} and separable to kinetic and cosmological terms.
It is written by
introducing an auxiliary field $\l(z)$,

\bea
       Z&=&\int D\l\tilde{Z}(\l), \\
      \tilde{Z}(\l)&=&\int Dg_{\m\n}D\o Df_i D\hi \ee{-S},
             \label{eq:a2}
\eea
\noindent where the action is written by
the kinetic part $S_{0}$ and the 'cosmological' term
$S_{\rm cos}$, $S=S_0+S_{\rm cos}$ and
\bea
       S_0&=&-{1 \over 4\pi}\int\,d^2z\sqrt{g}
            \bigl[ \hi R+4(\nabla\o)^2
                  -{1 \over 2}\sum_{i=1}^N(\nabla f_i)^2\bigr],\\
       S_{\rm cos}&=&-{1 \over 4\pi}\m^2\int\,d^2z\sqrt{g}
            \bigl[ (\l+4)\o^2-\l\hi\bigr].
                                         \label{eq:a3}
\eea
\noindent We can see the equivalency of this model and \rf{eq:a1}
at the classical level
by integrating over $\l$ and $\hi$ in (2) and by interpreting
$\o$ as $\ee{-\ph}$. So it is trivial to see that the classical properties
of this model are the same with that of the original DG \rf{eq:a1}.
Although it is not a
simple problem to quantize DG in the form \rf{eq:a1},
we can proceed a systematic procedure of quantization in our
equivalent model, as shown below.

It can be seen that $S_0$ is written
by the kinetic terms only in the conformal gauge
and its fully quantized form can be
obtained easily as shown below. It should be noticed that
$(S=)S_0$ is essentially equivalent to the model of
constant curvature, $R=0$ in this case, which is proposed
\cite{cham} as a
non-critical string model for arbitrary dimensions.
However our model is different from the constant curvature model
because of the existence of $S_{\rm cos}$. The procedure of the
quantization for $\m^2\neq 0$ can be performed
by treating $S_{\rm cos}$ as a small
perturbation according to the method of \cite{amb}.

Our strategy is as follows. First, obtain
the effective action in the form,

\beq
      \tilde{Z}(\l)= \ee{-S_{\rm eff}},
             \label{eq:a21}
\eeq
then examine the properties of the quantized theory by integrating out
$\l$ in (2) in the final step.
In ${\tilde Z}(\l)$,
$\l$ is treated as an external field. Here we take
the conformal gauge, $g_{\m\n}=\ee{2\r}\hg_{\m\n}$ with a fiducial
metric $\hg_{\m\n}$.
Due to the metric invariance, it is well known
that to obtain the fully quantized action is equivalent to getting
the one which is conformally invariant with respect to the
fiducial metric \cite{ddk}. According to this idea, the
effective action is obtained for $\m^2=0$ by adopting the following definition
of the norms for
each variables, $X^a=(\r, \hi, \o, f_i)$
where $a=0, 1, \cdots, 2+N$,

\beq
  \|\d X^a\|^2=\int\,d^2z\sqrt{\hg} (\d X^a)^2.
\label{eq:a4}
\eeq

\noindent By dividing ${\tilde Z}$ by the gauge volume,
the result is given by the following non-linear $\s$-model form,

\bea
     {\tilde Z}_q|_{\m^2=0}&=&\int D_{\hg}\r D_{\hg}\o D_{\hg}f_i D_{\hg}\hi
                        \ee{-S_{\rm eff}^{(0)}},\\
      S_{\rm eff}^{(0)}&=&{1 \over 4\pi}\int\,d^2z\sqrt{\hg}
            \left[ {1 \over 2}G_{ab}^{(0)}(X)\hg^{\a\b}
              \partial_\a X^a\partial_\b X^b
           +{\hat R}\Phi^{(0)}(X)+T^{(0)}(X) \right], \label{eq:a5}
\eea
and

\beq
   G_{ab}^{(0)}=\pmatrix{\k & -2 &  &  \cr
                         -2 &  0 &  &  \cr
                            &    & -8 &  \cr
                            &    &   & {\bf 1} \cr}, \  \  {}
   \Phi^{(0)}={1 \over 2}\k\r-\hi,
             \ \  T^{(0)}=0, \label{eq:a6}
\eeq
\noindent where ${\bf 1}$ in $G_{ab}^{(0)}$
denotes the unit matrix of dimension N and
\beq
       \k={23-N \over 3}. \label{eq:a7}
\eeq

As for the effective action for
$\m^2\neq 0$, it can be obtained perturbatively by expanding $S_{\rm eff}$
in the power series of $\m^2$ as follows,

\bea
  S_{\rm eff}&=&S_{\rm eff}^{(0)}+\m^2 S_{\rm eff}^{(1)}
                           + \m^4 S_{\rm eff}^{(2)}+\cdots  \\
             &=&{1 \over 4\pi}\int\,d^2z\sqrt{\hg}
            \left[ {1 \over 2}G_{ab}(X,\l)\hg^{\a\b}
              \partial_\a X^a\partial_\b X^b
           +{\hat R}\Phi(X,\l)+T(X,\l) \right]. \label{eq:a8}
\eea
where
\bea
G_{ab}&=& G_{ab}^{(0)}+\m^4G_{ab}^{(2)}+\cdots ,
           \  \  \Phi=\Phi^{(0)}+\m^4\Phi^{(2)}+\cdots,
           \nonumber \\
     & & \ \ {} T=T^{(0)} +\m^2 T^{(1)}+\cdots. \label{eq:a9}
\eea
It can be seen from the construction that $S_{\rm eff}^{(1)}$ is
the so-called
dressed term of $S_{\rm cos}$ and the terms of order $0(\m^2)$
do not appear in $G_{ab}$ and $\Phi$ as seen below.
The higher order terms are obtained by solving the equations of
motion derived from the following target space action,

\beq
S_t(\l)=
{1 \over 4\pi}\int\,d^dX\sqrt{G} e^{-2\Phi}
            \bigl[ R-4(\nabla\Phi)^2+
                  {1 \over 16}(\nabla T)^2+{1 \over 16}v(T)-\kappa
            \bigr],
\label{eq:a10}
\eeq

\noindent where $d=3+N$ and
the higher derivative terms are suppressed, and
\beq
v(T)=-2T^2+{1 \over 6}T^3+\cdots \  \ . \  \label{eq:a11}
\eeq
$\nabla_{\m}$ denotes the covariant derivative with
respect to the metric $G_{ab}$. We notice here that
$\l$ is not a coordinate but only a parameter.
And the equations to be solved are obtained as follows,

\bea
\nabla^2T-2\nabla\Phi \nabla T&=&{1 \over 2}v'(T),\label{eq:n4} \\
 \nabla^2\Phi-2(\nabla\Phi)^2&=&-{\kappa \over 2}
                       +{1 \over 32}v(T), \label{eq:n5} \\
 R_{ab}-{1 \over 2}G_{ab}R=
            -2\nabla_{a}\nabla_{b} \Phi &+&G_{ab}\nabla^2\Phi
            +{1 \over 16}\nabla_{a}T\nabla_{b}T-{1 \over 32}G_{ab}
            (\nabla T)^2. \label{eq:n6}
\eea
\noindent where $v'=dv/dT$.

\section{Quantum corrections}

We firstly determine $S_{\rm eff}^{(1)}$
or $T^{(1)}$ as the dressed operator of $S_{\rm cos}$
by using $S_{\rm eff}^{(0)}$ according the idea of \cite{nak}.
Two terms are contained in
$S_{\rm cos}$, the terms proportional to $\o^2$ and $\hi$, and
their dressed forms are given separately. The method is as follows.
Except for the prefactor $\m^2(\l +4)/4\pi$, the dressed term of
$\ee{2\r}\o^n$ is given as the coefficient of $\b^n/n!$ of the marginal
operator of
\[{\hat T}=\ee{\a\r+\b\o}\]
by expanding it in terms of $\b$. The
parameter $\a$ can be obtained by solving the lowest order equation
of (17) with respect to $\m^2$, and it is written as,

\beq
   {G^{(0)}}^{ab}[\partial_a\partial_b-2\partial_a\Phi^{(0)}\partial_b
                  +2]{\hat T}=0. \label{eq:b1}
\eeq
This equation gives,
\beq
   \a=-{1 \over 8}\b^2+2, \label{eq:b2}
\eeq
Then the dressed form of $\ee{2\r}\o^2$ is obtained as,
\beq
      \ee{2\r}(\o^2-{1 \over 4}\r). \label{eq:b3}
\eeq
Similarly we can get the dressed one for $\ee{2\r}\hi$
by solving \rf{eq:b1}, where ${\hat T}$ is replaced by
${\hat T}=\ee{\a\r+\b\hi}$. The result is as follows,
\beq
      \ee{2\r}(\hi-2\r). \label{eq:b4}
\eeq
As a result, we obtain
\beq
       S_{\rm eff}^{(1)}=-{1 \over 4\pi}\int\,d^2z\sqrt{\hg}\ee{2\r}
            \bigl[ (\l+4)(\o^2-{1 \over 4}\r)
                    -\l(\hi-2\r)\bigr].
                                         \label{eq:b5}
\eeq

The next task is to solve eqs.(18) and (19). It
should be noted that the lowest order ($O(1)$)
equation of (18),

\beq
 {G^{(0)}}^{ab}(\partial_a\partial_b\Phi^{(0)}
          -2\partial_aT^{(0)}\partial_bT^{(0)})=-{\k \over 2}
               ,\label{eq:b6}
\eeq

\noindent provides $\Phi^{(0)}$ which is given in (10). The
next order begins from $O(\m^4)$ which is needed to balance with
the substituted $T^{(1)}$ in eqs.(18) and (19).
Then the equations for $G_{ab}^{(2)}$ and $\Phi^{(2)}$
can be written down, but they include too many terms.
So we can solve them by restricting the space of
variables to that of $\r$, $\hi$ and $\o$ since $f_i$ decouple from
them. And we finally obtain seven equations by which we could solve seven
unknown functions $G_{ab}^{(2)}$ ($a,b=0,1,2$)
and $\Phi^{(2)}$. But the equations
contain second derivative terms, so we need many boundary conditions
to obtain a meaningful solution from them. From those solutions,
various counter terms needed for the theory can be seen and they are
necessary to solve the quantum mechanical black hole solutions
up to the order $O(\m^4)$. However our purpose is to examine
the properties of the theory up to $O(\m^2)$, so we do not need
their explicit form which will be given elsewhere.

Instead of giving the higher order counter terms of $\m^2$,
we examine here the quantum effects of DG
on the matter field by adding an interaction of a matter field
(with a coupling constant $\g$).
We restrict our attention on this gravitational
effect on the renormalization group property
of the matter fields. It is seen as follows by solving (18) and
(19) up to $\g^2$ by assuming $\m^2<<\g$.

Define $f_i$ as $\{f_j$ ($j=1\sim N-1$),
$\vp=f_N\}$, and add the following interaction term of $\vp$,

\beq
  S_{\rm int}={1 \over 4\pi}\g\int\,\sqrt{g}V(\vp)
               \label{eq:b7}
\eeq
to the original action $S$. Here we consider the Sine-Gordon model
as an example,
\beq
  V(\vp)=cos(p\vp),  \label{eq:b8}
\eeq
which has been previously examined in the usual 2d quantum gravity
without dilaton. Then we can see the difference between DG and the usual
2d gravity
by comparing our results obtained here and the one given in the previous
work \cite{amb}.

In order to simplify the problem, the fields $X^a$ in (10) are changed
as follows,
\beq
 \r={1 \over \sqrt{\k}}(\r '+\sqrt{2}\k '), \ \
      \hi=\sqrt{{\k \over 2}}\hi ', \ \ \ \o={1 \over 2\sqrt{2}}\o '
                       \label{eq:ch}.
\eeq
Then the form of $S_{\rm eff}^{(0)}$, which can be obtained in the limit of
$\g=\m^2=0$, can be written in the same form with \rf{eq:a5} but with the
different ${X'}^a=(\r ', \hi ', \o ', f_j, \vp)$, $G_{ab}^{(0)}$ and
$\Phi^{(0)}$ which are given as,

\beq
   G_{ab}^{(0)}=\pmatrix{1 & &  &  \cr
                          &  -1 &  &  \cr
                            &    & -1 &  \cr
                            &    &   & {\bf 1} \cr}, \  \  {}
   \Phi^{(0)}=\sqrt{{\k \over 2}}\r '.
             \ \  T^{(0)}=0, \label{eq:b9}
\eeq
In the following we suppress the prime of $X^a$ for the sake of the
brevity.

The calculations are performed according to the
procedure given above. But we are considering here
the perturbation in $\g$ instead of $\m^2$ under the assumption,
$1>>\g>>\m^2$. Firstly, we obtain the dressed operator ${\hat V}$
by solving the lowest equation of (17). It should be solved by taking
the following form
\beq
   {\hat V}(\vp)=V(\vp)\ee{\d(\r+\sqrt{2}\hi)}
                =\cos(p\vp)\ee{\d(\r+\sqrt{2}\hi)}. \label{eq:b10}
\eeq
Here we should notice that the exponent of the
dressed factor is changed due to
the shift, which is given in \rf{eq:ch}, of the original variable $\r$.
Equation (17) provides the following result,
\beq
   \d={2-p^2 \over \sqrt{\k}}. \label{eq:b11}
\eeq
{}From this we find a possible solution,
\beq
  p=\sqrt{2}, \ \ \, \d=0,   \label{eq:b12}
\eeq
which is corresponding to the Kosterlitz-Thouless fixed point.

Nextly, we solve the equations of $O(\g^2)$ of (18) and (19) by
substituting the solution \rf{eq:b10} and \rf{eq:b11}. They are solved
under the ansatz; (i) The functional coordinate space $\{X^a\}$ is
restricted to $\{\r, \hi, \vp\}$ since ${\hat V}$ is in this space.
(ii) Since the properties of the renormalization group equations are
determined by the $\r$-dependence of $G_{ab}$ of matter fields \cite{amb},
we solve the equations by taking the ansatz;
$G_{\vp\vp}^{(2)}=h(\r)$ and other $G_{ab}^{(2)}=0$. Then we obtain
the next seven equations,

\bea
 (\partial^2-2\sqrt{\k}\partial_{\r})\Phi^{(2)}
   -{\sqrt{\k} \over 4}\pa_{\r}h&=&-{1 \over 16}{\hat V}^2, \label{eq:b13} \\
    \pa_{a}\pa_{b} \Phi^{(2)}
         +{1 \over 4}[\d_a^0\d_b^0\pa_{\r}^2
                    +\d_a^1\d_b^1(\pa_{\r}^2+\sqrt{\k}\pa_{\r})]h
                      &=&{1 \over 32}\pa_a{\hat V}\pa_{\hat V}
                          , \label{eq:b14}
\eea

where $\pa^2=\pa_{\r}^2-\pa_{\hi}^2+\pa_{\vp}^2$ $=$
$\pa^2=\pa_{0}^2-\pa_{1}^2+\pa_{3+N}^2$. They are solved as follows,

\beq
       \Phi^{(2)}={1 \over 128}{\hat V}^2, \ \ \,
        h={p^2 \over 16\sqrt{\k}}\r. \label{eq:b15}
\eeq

This solution for $h(\r)$ is same with the one given in \cite{amb}, then the
same form of the renormalization group equations of $\d$
and $p$ are obtained by considering them
near the Kosterlitz-Thouless fixed point, $\g=0, p=\sqrt{2}$.
In fact, by parametrizing those parameters as, $p=\sqrt{2}+\ep$,
the following $\b$-functions are obtained,

\beq
   \b_{\ep}={\sqrt{2} \over 16}{2 \over \eta Q}\g^2, \ \ \,
   \b_{\g}=2\sqrt{2}{2 \over \eta Q}\ep \g,    \label{eq:b16},
\eeq

\noindent for the shift of the scale,
$\r\rightarrow \r+2dl/\eta$, where $\eta=2/Q$
\footnote{This is obtained by solving eq.(20) for
${\hat T}=\ee{\eta\r}$ in terms of (29).}
is defined by the dressed
factor of the cosmological constant, $\ee{\eta\r}$. The common factor
${2 \over Q\eta}$ in \rf{eq:b16} represent the quantum gravitational
effect on the matter $\b$-functions. The difference from the results
obtained in the previous work is the factor $\eta$, which characterizes
the 2d gravitational theory.

Finally, we comment on the string susceptibility ($\g_s$) of DG for
small $\m^2$. Neglecting the terms $O(\m^2)$, we obtain $\g_s$
according to \cite{ddk},

\beq
  \g_s=2-{\k \over 2\eta}\tilde{\hi}, \ \ \,
  \tilde{\hi}={1 \over 4\pi}\int d^2z\sqrt{\hg}\hat{R}, \label{eq:sus}
\eeq
where $\tilde{\hi}$ is the Euler character. Here we must be careful with
$\eta$ which is defined through the dressed factor $\ee{\eta\r}$
in terms of the original variable $\r$ (not $\r'$). Then we must
take $\eta =2$, which is obtained from eqs.(20) and (10) with
${\hat T}=\ee{\eta\r}$, in \rf{eq:sus}, and we obtain
\[ \g_s=2-{23-N \over 12}\tilde{\hi}.\]
This means that $\g_s$ is always real for any $N$. This is because
of the strong constraint on the curvature $R$. In fact,
$R=0$ for $\m^2=0$. But higher order corrections of $\m^2$ would
modify the above $\g_s$, and some constraint on $N$ may appear.
On this point, we will discuss elsewhere.

\section{Black holes and static solution}

Now we turn to the quantized dilaton gravity which includes $O(\m^2)$
correction,

\[ S_{\rm eff}=S_{\rm eff}^{(0)}+\m^2S_{\rm eff}^{(1)}. \]

\noindent And the matter interaction
term is suppressed here, e.g. $\g=0$.
After integrating over $\hi$ and $\l$, we obtain the next equations
of motion for $\r$ and $\o$,

\bea
   \pa_+\pa_-\o^2-{\k-7 \over 2}\pa_+\pa_-\r&=&
                     -\m^2\ee{2\r}(\o^2-{2\r+1 \over 8}), \label{eq:c1} \\
   {1 \over \o}\pa_+\pa_-\o+{1 \over 2}\pa_+\pa_-\r&=&
             -{1 \over 4}\m^2\ee{2\r}, \label{eq:c2}
\eea

\noindent where we have introduced
the light cone variables, $z^{\pm}=z^0\pm z^1$.
It is worthwhile to find an exact solution of the above equation, but
the above equations are approximate since the higher order terms of $O(\m^4)$
are suppressed. Then it is enough to find a solution accurate up to
$O(\m^2)$ for our purpose of examining
(i) the possibility of the black hole solutions and (ii) the stability
of static solutions.

Firstly, we study the possible black hole solution according to the
following ansatz,

\beq
  \ee{-2\r}=M-\m^2f(z), \  \,
  \o^2=m-\m^2g^{(1)}(z)+\m^4g^{(2)}(z)+O(\m^6), \label{eq:c3}
\eeq
where $M,m$ are constants. The classical equations of motion
derived from (4) and (5) give the following black hole
solution in the conformal gauge,
\beq
  \ee{-2\r}=\o^2={M_{\rm bh} \over \m^2}-\m^2z^+z^-, \label{eq:c5}
\eeq
where $M_{\rm bh}$ denotes the black hole mass, which is an arbitrary
constant. We had taken our ansatz so that the form of
the classical solution for $\r$, which provides the black hole
configuration,
is preserved if we obtain $f\propto z^+z^-$ and the higher order
corrections affect only on the configuration of $\o$ (the dilaton).

After a calculation, we can get the following solution of eqs.(38)
and (39),

\bea
    f(z)&=&f_0+f_1z^+z^-, \ \ \,
         g^{(1)}(z)=g_1+{m(1+f_1) \over 2M}z^+z^-, \label{eq:c6} \\
    g^{(2)}(z)&=&A{z^+}^2{z^-}^2+Bz^+z^-+C, \label{eq:c7}
\eea

\noindent where $f_0$, $g_1$ and C are arbitrary constants. $A$ and $B$ are the
calculable constants depending on $f_0$ and $g_1$, and
\beq
  f_1={2 \over m+\k-7}(m+{1 \over 4}\ln{M \over e}).
\eeq
\noindent Since the
scalar curvature is represented as $R=8\ee{-2\r}\pa_+\pa_-\r$, the
above black hole solution represents the same
$z$-dependence with the classical one except for
the normalization.
This means that we can find a black hole solution even in fully quantized
DG. It would be an interesting problem to consider the
relation between the result obtained here and an exact
2d black hole solution which was indicated by WZW models \cite{dij}.

Nextly, we turn to the problem of the the stability of the
static solution on which we should consider various problems.
To see the stability, we study the ADM mass
for the state which approaches to the linear dilaton vacuum,
$\r=0,\ \o=\ee{\m\s}$, at the spacial infinity,
$\s=(z^+-z^-)/2 \rightarrow \infty$. Such a configuration would be
written as follows,
\beq
    \r=\d\r,\ \ \, -\ln\o=-\m\s+\d\phi,      \label{eq:c8}
\eeq
at large $\s$, where $\d\r$ and $\d\phi$ decrease rapidly with $\s$.
By substituting \rf{eq:c8} into \rf{eq:c1} and \rf{eq:c2}, we obtain
the following linearized equations for $\d\r$ and $\d\phi$,
\bea
  \d\r+{1 \over \m}(\pa+{1 \over 4\m}\pa^2)\d\phi&=&
                                {1 \over 16}\ee{-2\m\s}, \label{eq:c9} \\
  (\pa^2+4\m^2)\d\r-2(\pa^2+2\m\pa)\d\phi&=&
                                0, \label{eq:c10}
\eea
where $\pa$ denotes the derivative with respect to $\s$.
They are solved as follows,
\beq
  \d\r=({1 \over 16}+a+b\s)\ee{-2\m\s}, \ \ \,
  \d\phi=(a+b\s)\ee{-2\m\s}, \label{eq:c11}
\eeq
where $a,b$ are arbitrary constants. From this asymptotic solution,
the ADM mass are obtained as,
\bea
  M_{\rm ADM}&=&2\ee{2\m\s}(\m\d\r+\pa_+\d\phi-\pa_-\d\phi) \\
             &=&2[b+\m({1 \over 16}-a)-\m b\s]. \label{eq:c12}
\eea
Even if we take $b=0$ so that the ADM mass does not diverge at
$\s\rightarrow \infty$, $M_{\rm ADM}$ is unbounded from below
because $a$ is arbitrary. This disease is common to other quantized
dilaton theory, and it is fatal to study the end of the black hole
by considering a falling matter.

\section{Conclusion}

New quantization scheme has been proposed for the dilaton gravity
by reformulating the original
DG so that the action can be separated to the kinetic
and a perturbation. The quantized theory has the black hole solution
which is qualitatively same with the classical one. But the ADM mass
of the static solution, which approaches to the linear dilaton vacuum
at spacial infinity, is unbounded from below. This
might be a sign that some
nonperturbative quantum effect \cite{ale}
would be necessary to approach the problem
of the Hawking radiation and the end of the black holes.

The string susceptibility of DG for small $\m^2$ is shown to be real
for any number of the conformal matters. Although the consideration
of higher order corrections are necessary, it would be worthwhile
to consider this DG as a
non-critical string model for higher dimensions than two.
We also find a quantum effect of DG on the renormalization group
equations of the interacting matter fields. The results are
consistent with the previous results obtained in the gravitation
without dilaton field.

%%%%%%%%%%%%%%%%%%%%%%%%%%%  Ref %%%%%%%%%%%%%%%%%%%%%%%%%%%%%%%%%%%%%%%%%


\newpage
\begin{thebibliography}{99}

\bibitem{bil}A. Bilal and C. Callan, Nucl. Phys. B394(1993) 73.
\bibitem{alw}S.P. de Alwis, Phys. Rev. D46(1992)5429.
\bibitem{gid}S. Giddings, and A. Strominger, Phys. Rev. D47(1993) 2454.
\bibitem{tan}Y. Tanii, Phys. Lett. B302(1993) 191.
\bibitem{cal}C. Callan, S. Giddings, J. Harvey and A. Strominger,
      Phys. Rev. D45 (1992) R1005.
\bibitem{ban}T. Banks, A. Dabholkar, M.R. Douglas and M. O'Loughlin,
           Phys. Rev. D45(1992) 3607.
\bibitem{kpz}V. Knizhnik, A. Polyakov and A. Zamolodchikov, Mod.Phys.Lett
             A3 (1988) 819.
\bibitem{ddk} F. David, Mod.Phys.Lett. A3 (1988) 1651; J. Distler
              and H. Kawai, Nucl.Phys. B321 (1989) 509.
\bibitem{amb}J.Ambjorn and K.Ghoroku, Int. J. Mod. Phys. A9(1994)5689.
\bibitem{cham}A. Chamseddine, Phys. Lett. B256(1991)379;Nucl. Phys.
             B368(1992)98.
\bibitem{nak}H. Kawai and R. Nakayama, Phys. Lett. B306(1993) 224.
\bibitem{dij}R. Dijkgrapff, H. Verlinde and E. Verlinde,
      Nucl. Phys. B371(1992) 269.
\bibitem{ale}G. Alejandro, F.D. Mazzitelli and C. Nunez,
       Preprint, Feb. 1995, hep-th/9411102.

%%%%%%%%%%%%%%%%%%%%%%%%%%%%%%%%%%%%%%%%%%%%%%%%%%%%%%%%%%%%%%%%%%%%%%%%%%%%
\end{thebibliography}
\end{document}